\documentclass[journal]{IEEEtran}
\ifCLASSINFOpdf
   \usepackage[pdftex]{graphicx}
\else
\fi
\usepackage{amsmath}
\usepackage{cite}

\usepackage{multirow}
\usepackage{multicol}
\usepackage{mathtools}

\usepackage{amsmath,epsfig}
\usepackage{amsfonts}
\usepackage{url}

\begin{document}
%
\title{Speckle2Void: Deep Self-Supervised SAR Despeckling with Blind-Spot Convolutional Neural Networks}
%
%
%

\author{Andrea~Bordone~Molini,
        Diego~Valsesia,
        Giulia~Fracastoro,
        and~Enrico~Magli
\thanks{The authors are with Politecnico di Torino -- Department of Electronics and Telecommunications, Italy. email: \{name.surname\}@polito.it.}}

%
%

\markboth{}%
{Bordone Molini \MakeLowercase{\textit{et al.}}: Speckle2Void: Deep Self-Supervised SAR Despeckling with Blind-Spot Convolutional Neural Networks}
%



\maketitle

\begin{abstract}
Information extraction from synthetic aperture radar (SAR) images is heavily impaired by speckle noise, hence despeckling is a crucial preliminary step in scene analysis algorithms. The recent success of deep learning envisions a new generation of despeckling techniques that could outperform classical model-based methods. However, current deep learning approaches to despeckling require supervision for training, whereas clean SAR images are impossible to obtain. In the literature, this issue is tackled by resorting to either synthetically speckled optical images, which exhibit different properties with respect to true SAR images, or multi-temporal SAR images, which are difficult to acquire or fuse accurately. In this paper, inspired by recent works on blind-spot denoising networks, we propose a self-supervised Bayesian despeckling method. The proposed method is trained employing only noisy SAR images and can therefore learn features of real SAR images rather than synthetic data. Experiments show that the performance of the proposed approach is very close to the supervised training approach on synthetic data and superior on real data in both quantitative and visual assessments.

\end{abstract}


%
\IEEEpeerreviewmaketitle

\section{Introduction}

Synthetic Aperture Radar (SAR) is a coherent imaging system and as such it strongly suffers from the presence of speckle, a signal dependent granular noise. Speckle noise makes SAR images difficult to interpret, preventing the effectiveness of scene analysis algorithms for, e.g., image segmentation, detection and recognition.
Several despeckling methods applied to SAR images have been proposed working either in spatial or transform domain. 
The first attempts at despeckling employed filtering-based techniques operating in spatial domain such as Lee filter \cite{LEE198124}, Frost filter \cite{4767223}, Kuan filter \cite{1165131}, and Gamma-MAP filter \cite{doi:10.1080/01431169308953999}.
Wavelet-based methods \cite{1105905,1166595} enabled multi-resolution analysis. More recently, non-local filtering methods attempted to exploit self-similarities and contextual information. A combination of non-local approach, wavelet domain shrinkage and Wiener filtering in a two-step process led to SAR-BM3D \cite{5989862}, a SAR-oriented version of BM3D \cite{4271520}.

In recent years, deep learning techniques have become the benchmark in many image processing tasks, achieving exceptional results in problems such as image restoration \cite{7839189}, super resolution \cite{deepsum}, semantic segmentation \cite{7298965}, and many more.
Recently, some despeckling methods based on convolutional neural networks (CNNs) have been proposed \cite{8053792,DilatedSAR}, attempting to leverage the feature learning capabilities of CNNs. Such methods use a supervised training approach where the network weights are optimized by minimizing a distance metric between noisy inputs and clean targets. 
However, clean SAR images do not exist and supervised training methods resort to synthetic datasets where optical images are used as ground truth and their artificially speckled version as noisy inputs. This creates a domain gap between the features of synthetic training data and those of real SAR images, possibly leading to the presence of artifacts or poor preservation of radiometric features when despeckling real SAR images.
SAR-CNN \cite{CozzolinoCNN} addressed this problem by averaging multi-temporal SAR data of the same scene in order to obtain an approximate (finite number of looks) ground truth. However, acquisition of multi-temporal data, scene registration and robustness to temporal variations can be challenging, leading to a sub-optimal rejection of speckle.

Recently, self-supervised denoising methods \cite{pmlr-v80-lehtinen18a,Krull2018Noise2VoidL,batson2019noise2self,laine2019high} proved, under certain assumptions, to be a valid alternative when it is not possible to have access to clean images. In particular,  the two methods in \cite{Krull2018Noise2VoidL,laine2019high} deal with a single noisy version of each image in the dataset. These two works make use of a modified version of the classical CNN, called blind-spot convolutional network, to reconstruct each clean pixel exclusively from its neighboring pixels. The target pixel itself is kept hidden by the blind spot operation during training in order to prevent the network from learning the identity mapping and just copying the noisy pixel in the final denoised image.
Self-supervision thus allows to exploit the potential of deep learning in those fields where the ground truth is not accessible, such as SAR imaging.

Inspired by these works, in this paper we present Speckle2Void, a self-supervised Bayesian despeckling framework that enables direct training on real SAR images. Our method bypasses the problem of training a CNN on synthetically-speckled optical images, thus avoiding any domain gap and enabling learning of features from real SAR images. It also avoids the inherent difficulty in constructing multitemporal datasets, as done in \cite{CozzolinoCNN}. Our main contributions can be summarized as follows:
\begin{itemize}
    \item we formulate a Bayesian model to characterize the speckle and the prior distribution of pixels in the clean SAR image, conditioned on their neighborhoods;
    \item we propose an improved version of the blind-spot CNN architecture in \cite{laine2019high} and a regularized training procedure with a variable blind-spot shape in order to account for the autocorrelation of the speckle process;
    \item we present two versions of Speckle2Void: a local version with classical convolutional layers and a non-local version to incorporate information from both spatially-neighboring as well as distant pixels to exploit self-similarity, albeit at higher computational complexity;
    \item we achieve remarkable despeckling performance, showing how our self-supervised approach is better than model-based techniques, close to the deep learning methods requiring supervised training on synthetic images and superior to them on real SAR data.
\end{itemize}

A preliminary version of this work appeared in \cite{Bordone_sar}, showing the basic principles of the proposed approach. This paper significantly expands the treatment with improvements on network modeling, on the loss function and on the training procedure. In particular, it solves the problem of the residual granularity in the despeckled images in \cite{Bordone_sar}, by showing the importance of properly decorrelating the speckle process and carefully designing the blind-spot shape.

The remainder of this paper is organized as follows. Section \ref{sec:relatedWork} introduces related works on SAR despeckling. Section \ref{sec:background} provides the background knowledge on the Bayesian framework adopted in this work. Section \ref{sec:method} details the proposed statistical models and the regularized blind-spot network with variable structure. Section \ref{sec:results} contains results and performance evaluation. Section \ref{sec:conclusions} draws some conclusions.


\section{Related work}
\label{sec:relatedWork}

\subsection{SAR Despeckling}
The last decades have seen a multitude of SAR image despeckling methods, that can be broadly categorized into four main approaches: spatial-domain methods, wavelet-domain methods, non-local methods and deep learning methods.
Filtering-based techniques such as Lee filter \cite{LEE198124}, Frost filter \cite{4767223}, Kuan filter \cite{1165131} represent the early attempts to solve SAR despeckling and they operate in spatial domain.
Subsequent works in spatial domain aimed to reduce speckle under a non-stationary multiplicative speckle assumption. A popular example is represented by the Bayesian maximum a posteriori (MAP) approaches aiming to give a statistical description to the SAR image. A few MAP-based works have been proposed and the most representative is the $\Gamma$-MAP filter \cite{doi:10.1080/01431169308953999} that solves the MAP equation modeling both the radar reflectivity and the speckle noise with a Gamma distribution.

Wavelet-based methods proved to be more effective than spatial domain ones, enabling multi-resolution analysis and boosting analysis under non-stationary characteristics. They despeckle SAR images in the transform domain by estimating despeckled coefficients and then by applying the inverse transform to obtain the cleaned SAR image. 
A first subclass of wavelet based methods solve the despeckling problem with a homomorphic approach, consisting in applying a logarithmic transform of the data to convert the multiplicative noise into an additive one.  
The works in \cite{413278,10.1117/12.279681} applied the traditional wavelet shrinkage based on hard- and soft-thresholding with an empirical selection of the threshold. Further wavelet-based methods \cite{1221775,1288366,4162521,1673449} introduce prior knowledge about the log-transformed reflectance in the wavelet domain, employing a MAP estimator.
Most of the wavelet-based homomorphic approaches do not compensate for the bias in the reconstructed images resulting from the mean of the log-transform speckle. To cope with this problem, a non-homomorphic approach has been considered by some works \cite{1025027,1323120,892442,1709983} in the wavelet domain, dealing with a signal-dependent speckle whose distribution parameters are harder to be estimated.

In general, both spatial domain and wavelet domain techniques yield limited detail preservation with the introduction of severe artifacts. The amount of information provided by a local window is quite limited and the need of incorporating more information from the neighborhood led to the proliferation of non-local methods. 
The pioneering work in this field is represented by the non-local mean (NLM) filter \cite{nonlocalmeans} that performs a weighted average of all pixels in the image and the weights depend on their similarity with respect to the target pixel. The weights are defined by computing the Euclidean distance between a surrounding patch centered at a neighboring pixel and a local patch centered at the target pixel.
In \cite{5196737}, the Probabilistic Patch-Based (PPB) algorithm has been proposed to adapt the non-local means approach to SAR despeckling. The authors devised a patch similarity measure that generalizes to the case of multiplicative, non-Gaussian speckle.

NLM inspired a number of extensions in the Gaussian noise context such as the Block-Matching 3D (BM3D) algorithm \cite{4271520}, a combination of non-local approach, wavelet domain shrinkage and Wiener filtering in a two-step process. 
One of the most popular SAR despeckling algorithm is the SAR version of BM3D \cite{4271520} (SAR-BM3D) that follows the same BM3D phases with an adaptation to the SAR statistics in the grouping phase where the same PPB similarity measure is used. Moreover the hard-thresholding and Wiener filtering, suitable in the Gaussian noise context, are replaced with an LMMSE estimator (based on an additive signal-dependent noise model).

The success of deep learning on many tasks involving image processing has suggested that the powerful learning capabilities of CNNs could be exploited for SAR despeckling and a few works have started addressing the problem.
Chierchia et al. \cite{CozzolinoCNN} proposed SAR-CNN, which applies a DnCNN-like \cite{DnCnnZhang} supervised denoising approach to SAR data. They exploit the homomorphic approach to deal with multiplicative noise model and use a new similarity measure for speckle noise distribution as loss function rather than the usual Euclidean distance. Clean data for training are obtained by averaging multitemporal SAR images.
Wang et al. \cite{8053792} proposed a residual CNN (ID-CNN) trained on synthetic SAR images, to directly estimate the noise in the original domain, and, hence, the despeckled image is obtained by dividing the noisy image by the estimated noise. Training is once again supervised using synthetically speckled optical images and carried out with the Euclidean distance and a total variation regularization as loss function.
Several subsequent deep learning works \cite{8313133,DilatedSAR,Gui,HDRANet,ZhangJing2019SIDU,Lattari} proposed slight variations on the topic by introducing different architectures and losses, but all under the supervised training umbrella using synthetically speckled SAR images.
In \cite{8313133} the authors proposed IDGAN, a deep learning SAR despeckling method based on a generative adversarial network (GAN) and trained using a weighted combination of Euclidean loss, perceptual loss and adversarial loss.
In \cite{Gui}, a dilated densely connected network (SAR-DDCN) trained with Euclidian distance, was proposed to enlarge the receptive field and to improve feature propagation and reuse. 
A combination of hybrid dilated convolutions and both spatial and channel attention modules through a residual architecture called HDRANet was proposed in \cite{HDRANet}, to further improve the feature extraction capability. More recently, Cozzolino et al. \cite{NonlocalCozzolino} proposed a method that combines the classical non-local means method with the power of CNN, where NLM weights are assigned by a convolutional neural network with non local layers.

Until now, the power of CNN has not been fully exploited yet, since most of the works in literature make use of synthetic SAR images.
Inspired by the recent blind-spot CNN denoising works, we tackle SAR despeckling with a self-supervised Bayesian framework relying on blind-spot CNNs.

\subsection{Self-supervised denoising with CNNs}\label{sec:related_selfsup}

During the last year, significant advances have been made on deep learning approaches to denoising that do not require ground-truth, showing that it is possible to reach performance close to that exhibited by fully-supervised methods. These new self-supervised denoising methods have been developed on natural images, but it is quite clear that extending them to the SAR context is appealing, as significant speckle noise is always present in SAR acquisitions.
Noise2Noise \cite{pmlr-v80-lehtinen18a} proposed to use pairs of images with the same content but independent noise realizations. The main drawback of this method is the difficulty of accessing multiple versions of the same scene with independently drawn noise realizations. Yuan et al. \cite{Yuan2019BlindSI} presented a despeckling method based on the idea of Noise2Noise \cite{pmlr-v80-lehtinen18a}, but still simulating speckle on a dataset based on ImageNet.
Noise2Void \cite{Krull2018Noise2VoidL} and Noise2Self \cite{batson2019noise2self} further relax the constraints on the dataset, requiring only a single noisy version of the training images, by introducing the concept of blind-spot networks. Assuming spatially uncorrelated noise, and excluding the center pixel from the receptive field of the network, the network learns to predict the value of the center pixel from its receptive field by minimizing the $\ell_2$ distance between the prediction and the noisy value. The network is prevented from learning the identity mapping because the pixel to be predicted is removed from the receptive field. Notice that this is also the reason for the uncorrelated noise assumption. The blind-spot scheme used in Noise2Void \cite{Krull2018Noise2VoidL} is carried out by a simple masking method that hides one pixel at a time,  processing the entire image to learn to reconstruct a single cleaned pixel.
Laine et al. \cite{laine2019high} devised a novel blind-spot CNN architecture capable of processing the entire image at once, increasing the efficiency. They also introduced a Bayesian framework to include noise models and priors on the conditional distribution of the blind spot given the receptive field.

\section{Background}
\label{sec:background}

CNN denoising methods estimate the clean image by learning a function that takes each noisy pixel and combines its value with the local neighboring pixel values (receptive field) by means of multiple convolutional layers interleaved with non-linearities. Taking this from a statistical inference perspective, a CNN is a point estimator of $p(x_i|y_i,\Omega_{y_i})$, where $x_i$ is the $i^\text{th}$ clean pixel, $y_i$ is the $i^\text{th}$ noisy pixel and $\Omega_{y_i}$ represents the receptive field composed of the noisy neighboring pixels, excluding $y_i$ itself.
Noise2Void and Noise2Self predict the clean pixel $x_i$ by relying solely on the neighboring pixels and using $y_i$ as a noisy target. By doing so, the CNN learns to produce an estimate of $\mathbb{E}_{x_i}[x_i|\Omega_{y_i}]$, using the $\ell_2$ loss when in presence of Gaussian noise.
The drawback of these methods is that the value of the noisy pixel $y_i$ is never used to compute the clean estimate. 

The Bayesian framework devised by Laine et al. \cite{laine2019high} explicitly introduces the noise model $p(y_i|x_i)$ and conditional pixel prior given the receptive field $p(x_i|\Omega_{y_i})$ as follows: 
\begin{align*}
p(x_i|y_i,\Omega_{y_i}) \propto p(y_i|x_i) p(x_i|\Omega_{y_i}).
\end{align*}
The role of the CNN is to predict the parameters of the chosen prior $p(x_i|\Omega_{y_i})$. The denoised pixel is then obtained as the posterior mean (MMSE estimate), i.e., it seeks to find $\mathbb{E}_{x_i}[x_i|y_i,\Omega_{y_i}]$.
Under the assumption that the noise is pixel-wise i.i.d., the CNN is trained so that the data likelihood $p(y_i|\Omega_{y_i})$ for each pixel is maximized. The main difficulty involved with this technique is the definition of a suitable prior distribution that, when combined with the noise model, allows for closed-form posterior and likelihood distributions. We also remark that while imposing a handcrafted distribution as $p(x_i|\Omega_{y_i})$ may seem very limiting, it is actually not since i) that is the \textit{conditional} distribution given the receptive field rather than the raw pixel distribution, and ii) its parameters are predicted by a powerful CNN on a pixel-by-pixel basis.

\begin{figure*}[t]
\centering
\includegraphics[width=7.1in]{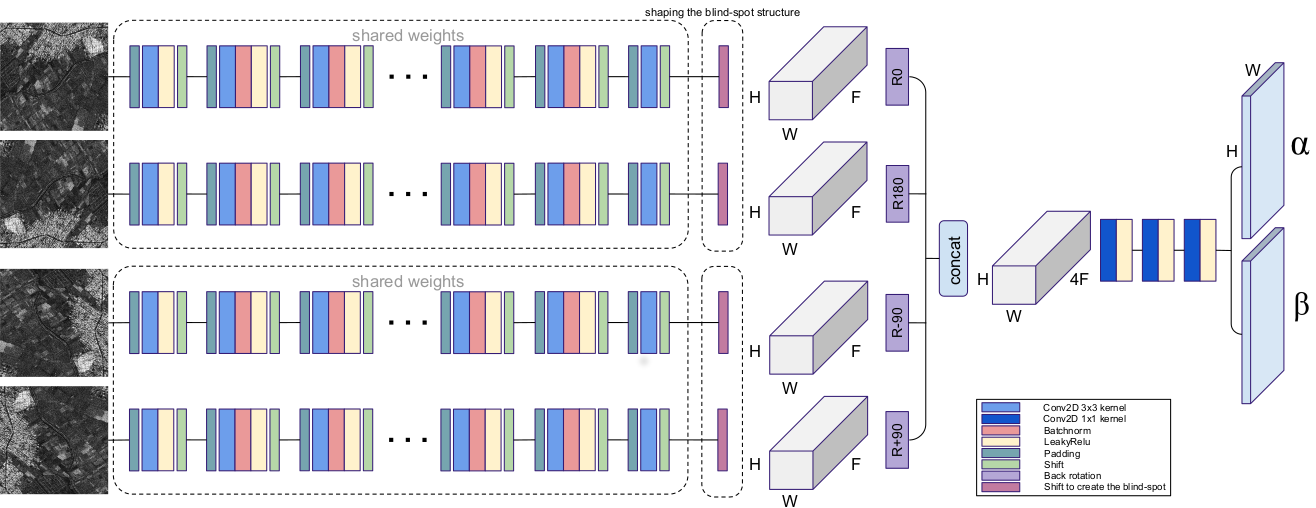}
\caption{Speckle2Void takes as input four rotated versions of an image. Each branch processes a specific rotation to compute the receptive field in a specific direction. Subsequently, the four half-plane receptive fields are shifted to achieve the desired blind-spot shape, rotated back and concatenated. As last, a series of 2D convolutions with kernel 1x1 are used to fuse the four receptive fields and generate the parameters of the inverse gamma for each pixel.}
\label{fig:Architecture}
\end{figure*}
 
\section{Proposed method}
\label{sec:method}

\begin{figure}[t]
\centering
\includegraphics[width=0.8\linewidth]{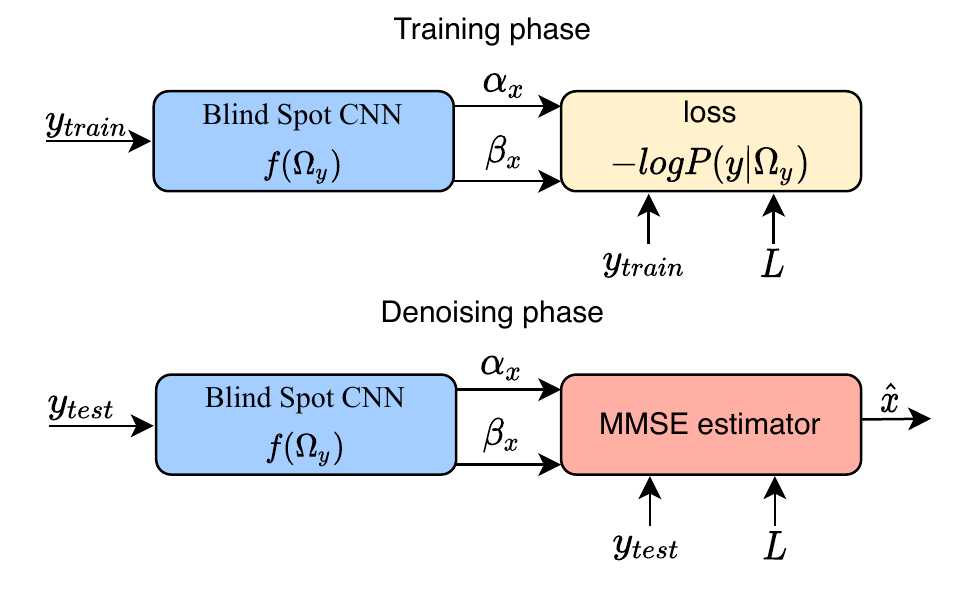}
\vspace{-0.4cm}
\caption{Scheme depicting the training and the testing phases. During training phase the blind-spot network is trained to minimize the negative log of the noisy data likelihood to estimate $\alpha_{x_i}$ and $\beta_{x_i}$ for each pixel. In testing phase, the MMSE estimator generates the final clean image, combining together the parameters of the pixel prior, the noisy pixel and the parameter of noise distribution.}
\vspace{-0.3cm}
\label{fig:summary}
\end{figure}

Following the notation in Sec. \ref{sec:background}, this section presents the Bayesian model we adopt for SAR despeckling, the training procedure and the blind-spot architecture. A summary is shown in Figs. \ref{fig:Architecture} and \ref{fig:summary}.

\subsection{Model}

We consider the multiplicative SAR speckle noise model: $y_i = n_i x_i$, where $x$ represents the unobserved clean image in intensity format and $n$ the spatially uncorrelated multiplicative speckle. Concerning noise modeling, one common assumption is that it follows a Gamma distribution with unit mean and variance $1/L$ for an $L$-look image and has the following probability density function: 
\begin{align*}
p(n)&=\frac{1}{\Gamma(L)} L^L n^{L-1} e^{Ln}
\end{align*}
where $\Gamma(.)$ denotes the Gamma function and $n \geq 0$, $L \geq 1$.
The aim of despeckling is to estimate intensity backscatter $x$ from the observed intensity return $y$.

We model the conditional prior distribution given the receptive field as an inverse Gamma distribution with shape $\alpha_{x_i}$ and scale $\beta_{x_i}$:
\begin{align*}
    p(x_i|\Omega_{y_i}) = \mathrm{inv}\Gamma(\alpha_{x_i},\beta_{x_i}),
\end{align*}
where $\alpha_{x_i}$ and $\beta_{x_i}$ depend on $\Omega_{y_i}$, since they are the outputs of the CNN at pixel $i$.
Assuming the noise to be Gamma-distributed, i.e., $n_i \sim \Gamma(L,L)$, then by the scaling property of the Gamma distribution, we obtain that $y_i|x_i \sim \Gamma(L,\frac{L}{x_i})$.
We can now write the unnormalized posterior distribution as:
\begin{align*}
p(x_i|y_i,\Omega_{y_i}) &\propto p(y_i|x_i) p(x_i|\Omega_{y_i}),\\
p(x_i|y_i,\Omega_{y_i}) &\propto \frac{1}{\Gamma(L)} \left(\frac{L}{x_i}\right)^L y^{L-1}_i e^{\frac{L}{x_i} y_i} \frac{\beta^{\alpha_{x_i}}_{x_i}}{\Gamma(\alpha_{x_i})} \frac{e^{\frac{\beta_{x_i}}{x_i}}}{x^{\alpha_{x_i}+1}},\\
&\propto \frac{e^{\frac{Ly_i+\beta_{x_i}}{x_i}}}{x^{\alpha_{x_i}+L+1}}
\end{align*}

For the chosen prior and noise models, the posterior distribution has still the form of an inverse Gamma:
\begin{align}\label{eq:posterior}
    p(x_i|y_i,\Omega_{y_i}) = \mathrm{inv}\Gamma(L+\alpha_{x_i},\beta_{x_i}+Ly_i).
\end{align}

Finally, the noisy data likelihood $p(y_i|\Omega_{y_i})$ can be obtained in closed form as:
\begin{align}\label{eq:likelihood}
p(y_i|\Omega_{y_i})&=\frac{L^L y^{L-1}_i}{\beta^{-\alpha_{x_i}}_{x_i} \mathrm{Beta}(L,\alpha_{x_i}) (\beta_{x_i} + L y_i)^{L+\alpha_{x_i}}},
\end{align}
with the Beta function defined as $\mathrm{Beta}(L, \alpha_{x_i}) = \frac{\Gamma(L)\Gamma(\alpha_{x_i})}{\Gamma(L+\alpha_{x_i})}$.
This distribution was first introduced in \cite{581981} to model the intensity return in SAR images and it is known as the $G^0_I$ distribution. According to \cite{581981}, the $G^0_I$ distribution is a very general model, able to accommodate extremely homogeneous areas as well as scenes such as urban areas.

\subsection{Training}\label{sec:training}
\label{subsec:training}

The training procedure learns the weights of the blind-spot CNN. The blind-spot CNN processes the noisy image to produce the estimates for parameters $\alpha_{x_i}$ and $\beta_{x_i}$ of the inverse gamma distribution $p(x_i|\Omega_{y_i})$ used as prior. 
It is trained to minimize the negative log likelihood $p(y_i|\Omega_{y_i})$ for each pixel, so that the estimates of $\alpha_{x_i}$ and $\beta_{x_i}$ fit the noisy observations. 

As stated in Sec.\ref{sec:related_selfsup}, training a blind-spot network requires noise to be spatially uncorrelated, so that the CNN is prevented from exploiting the latent correlation to reproduce the noise in the blind spot. While many works assume that SAR speckle is uncorrelated, the SAR acquisition and focusing system has a point spread function (PSF) that correlates the data. To cope with this, we apply a pre-processing whitening procedure, such as the one proposed by Lapini et al. \cite{6487399} to decorrelate the speckle. In \cite{6487399}, the authors use the complex SAR data after focusing to estimate the PSF of the system and approximately invert it, achieving the desired decorrelation and showing that this step boosts the performance of any despeckling algorithm relying on the uncorrelated speckle assumption. This whitening step is especially critical in the proposed approach due to the high capacity of neural networks to overfit even random patterns. 

However, perfect decorrelation is in practice impossible and the residual correlation could limit the performance of the blind-spot CNN. For this reason, we modify the basic design of the blind-spot CNN by Laine et al. \cite{laine2019high}, and introduce a variable-sized blind spot. If noise correlation cannot be removed by other means, one could consider the width of the autocorrelation function of the noise and set a blind spot that is wide enough to cover the peak of the autocorrelation. This ensures that the receptive field contains a negligible amount of information for the reproduction of the noise component of the pixel to be estimated. However, this inevitably reduces the amount of information that can be exploited by the CNN, as the content of the immediate neighbors of a pixel is the most similar to that of the pixel itself. Therefore, a larger blind spot trades off more effective noise suppression with a less accurate (appearing as blurry) prediction.  

To achieve a finer control about this trade-off, we devise a regularized training procedure that allows to tune the degree of reliance of the CNN on the immediate neighbors, leading to an improvement of the high frequency details in the denoised image, while still suppressing most of the noise correlation.
During training, we randomly alternate, with predefinied probabilities, a $1\times1$ blind spot and a larger blind spot that can have arbitrary shape to match the noise autocorrelation.
This mechanism allows the network weights to learn how to partially exploit the neighboring pixels belonging to the larger blind-spot but at the same time not to rely too much on them, in order to prevent from overfitting the noise components. 
During testing, a $1\times1$ blind spot is used, thus only excluding the center pixel, and exploiting the closest neighbors. Due to their weak training, these neighbors allow to recover some high frequency image content, which is the stronger signal present, while not being able to exploit the weaker correlations in the noise.

\subsection{Testing}

In testing, the blind-spot CNN processes the noisy SAR image to estimate $\alpha_{x_i}$ and $\beta_{x_i}$ for each pixel. 
The despeckled image is then obtained through the MMSE estimator, i.e., the expected value of the posterior distribution in Eq. \eqref{eq:posterior}, as:
\begin{align*}
    \hat{x}_i = \mathbb{E}[x_i|y_i,\Omega_{y_i}] = \frac{\beta_{x_i} + L y_i}{L+\alpha_{x_i}-1}.\\[-16pt]
\end{align*}
Notice that this estimator combines both the per-pixel prior estimated by the CNN and the noisy observation.

\subsection{Loss function}
As mentioned in Sec. \ref{sec:training}, the blind-spot CNN is trained by minimizing the negative log likelihood of the noisy observations, based on the estimated parameters $\alpha_{x_i}$ and $\beta_{x_i}$ of the prior.
Moreover, we incorporate a total variation (TV) component, computed over the posterior image, to further promote smoothness.
Our final loss function is as follows:
\begin{align*}
l &= - \sum_{i} \log p(y_{i}|\Omega_{y_{i}}) + \lambda_{TV} TV(\hat{x})
\end{align*}
where $p(y_{i}|\Omega_{y_{i}})$ is defined in Eq. \eqref{eq:likelihood}, the TV term is the anisotropic version of the total variation $TV(\hat{x})=\sum_{i,j} |\hat{x}_{i+1,j}-\hat{x}_{i,j}| + |\hat{x}_{i,j+1}-\hat{x}_{i,j}|$ and $\lambda_{TV}$ is a hyperparameter to tune the desired degree of smoothness.

\subsection{Blind-spot architecture}

 \begin{figure}[t]
\centering
\includegraphics[width=3.5in]{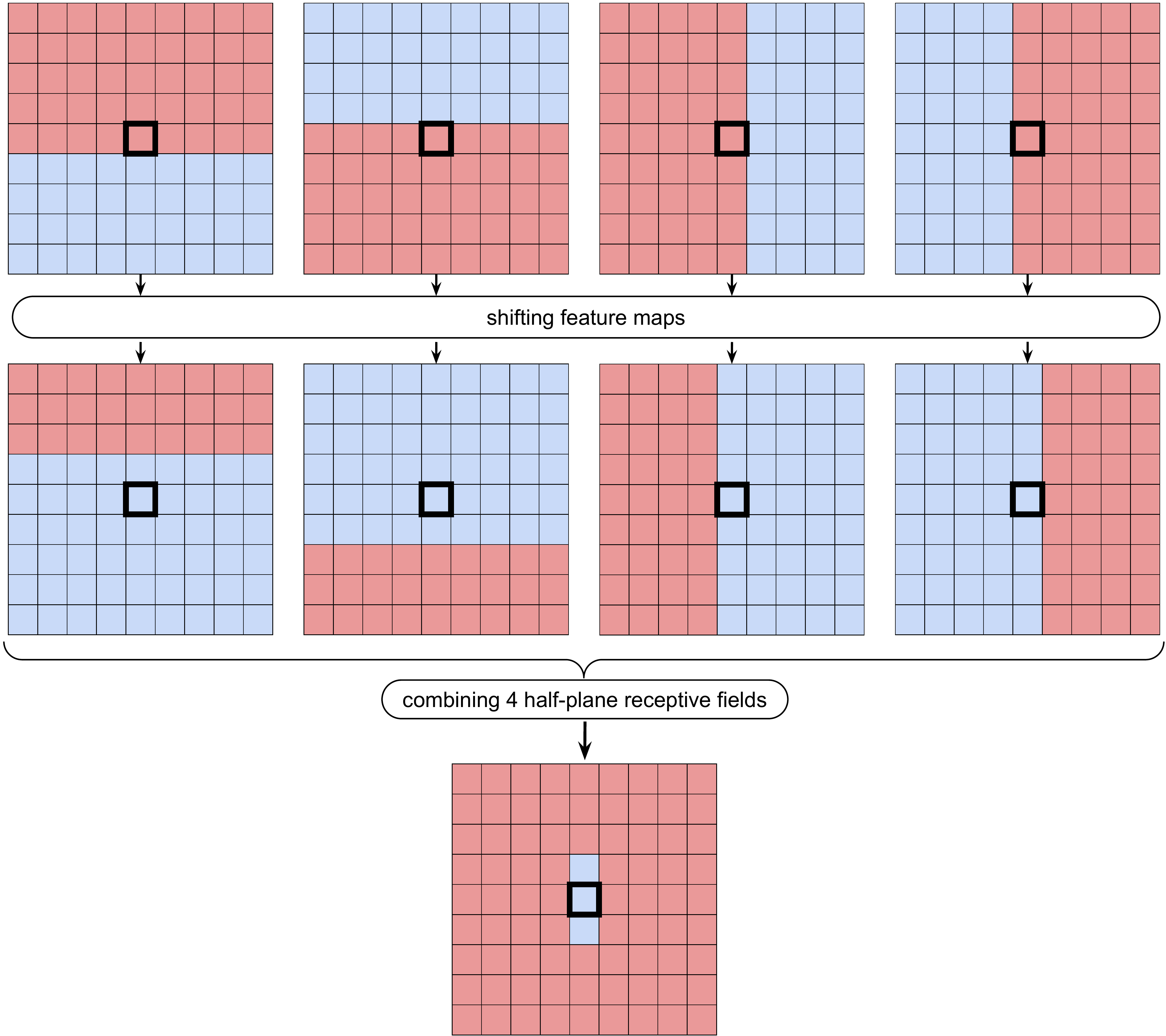}
\caption{Visual depiction of the operations performed by the blind-spot network to constrain the receptive field related to a center pixel to exclude the center pixel itself and two pixels in the vertical direction.}
\label{fig:blind_spot_3x1}
\end{figure}

The rationale behind the blind-spot network is to introduce a pixel-sized hole in the receptive field, in order to prevent the network from learning the identity mapping. 
Our model is built upon the architecture by Laine et al.\cite{laine2019high}, who designed a CNN architecture to naturally account for the blind spot in the receptive field, thus increasing training efficiency. They cleverly implemented shift and padding operations on the feature maps at each layer, in order to limit the receptive field to grow in a specific direction, excluding the center pixel from the computation. 
Their architecture is composed of four different CNNs, each responsible of limiting the receptive field to extend in a single direction by means of shift and padding operations on the feature maps at each layer. The four subnetworks produce four limited receptive fields that extend strictly above, below, leftward and rightward of the target pixel. In order to reduce the number of trainable parameters, they feed four rotated versions of each input image to a single network that computes the receptive field in a specific direction.
The four limited receptive fields are finally combined through a series of 2D convolutions with $1\times1$ filters, ensuring no further expansion of the receptive field.
To perform this particular computation, classical 2D convolutional layers are used but their receptive field is limited to grow in a direction by shifting the feature map in the opposite direction by an offset of $\lfloor{k/2}\rfloor$ pixels, where $k \times k$ is the kernel size, before performing the convolution operation.
At the end of the network, each of the four limited receptive fields still contains the center row/column, so the center pixel as well. To exclude it, the feature maps are shifted by one pixel before combining them.

An overview of the blind-spot network used by Speckle2Void is shown in Fig. \ref{fig:Architecture}.
Speckle2Void modifies the basic architecture by Laine et al.\cite{laine2019high} described above to allow more flexibility in shaping the blind-spot. In principle, if the final shift applied to each of the four directional receptive fields was different from one another, we would be able to control the size of the blind spot in each direction. In SAR images, the azimuth and range directions may exhibit different statistical properties, including the residual noise autocorrelation. We therefore account for that by only sharing weights between the two branches processing the receptive field oriented as the azimuth or range directions, instead of sharing them for all four branches as in \cite{laine2019high}. Furthermore, as shown in Fig. \ref{fig:blind_spot_3x1}, Speckle2Void can apply one shift in the azimuth direction and a different shift in the range one.

\subsection{Non local convolutional layer and its adaptation to blind-spot networks}

The blind-spot CNN used by Speckle2Void also comes in two versions. The ``local'' version of Speckle2Void is composed by a series of classic 2D convolutional layers, each followed by Batch normalization \cite{ioffe2015batch} and a Leaky-ReLU non-linearity.
The ``non-local'' version adds several non-local layers, as defined in \cite{liu2018non}. Non-local layers introduce a dynamic weighted function of the feature vectors that help retrieving more information from a wider image context. In particular, they allow to exploit non-local self-similarity, which can be effective in recovering the information hidden by the blind spot, without encountering the problem of noise correlation as it is drawn from spatially-distant areas. However, exploiting non-locality incurs a significant penalty in terms of computational cost.

The non-local module proposed by NLRN \cite{liu2018non} uses a soft block matching approach and applies the Euclidean distance with linearly embedded Gaussian kernel as distance metric. The non-local layer is designed to work in a traditional CNN architecture, and requires introducing a masking technique to adapt it to the blind-spot architecture used by Speckle2Void. 
In \cite{liu2018non}, the linear embeddings are defined as follows:
\begin{align*}
\Phi(X_{ij}) \hspace{-1pt} &= \hspace{-1pt} \phi(X_{ij},X_{p_{ij}}) = \text{exp}\{\theta(X_{ij})\psi(X_{p_{ij}}))\}, \forall i,j , \\
\theta(X_{ij}) \hspace{-1pt} &= \hspace{-1pt} X_{ij}W_{\theta}, \psi(X_{p_{ij}}) \hspace{-1pt} = \hspace{-1pt} X_{p_{ij}}W_{\psi}, G(X_{ij}) \hspace{-1pt} = \hspace{-1pt} X_{p_{ij}}W_{g}, \forall i,j.  
\end{align*}
$\Phi(X_{ij})$ represents the distance metric to encode the non local correlation between the feature vector in position $i,j$ and each neighbours in the patch $X_{p_{ij}}$. $\Phi(X_{ij})$ has shape $1 \times q \times q$ where $q \times q$ denotes the spatial size of the neighbour patch centered at pixel $i,j$. 
$\theta(X_{ij})$ represents the embedding associated to the feature vector in position $i,j$ with shape $1 \times l$ where $l$ is the number of features. $\psi(X_{p_{ij}})$ represents the embeddings associated to each feature vector in the neighbour patch $p$ centered at $i,j$ with shape $q \times q \times m$ where $m$ is the number of features.
The transformation weights $W_{\theta},W_{\psi},W_{g}$ used to compute the embeddings have shape $m \times l$, $m \times l$, $m \times m$ respectively, and are trainable weights.
We add a masking operation to the non-local layer proposed in \cite{liu2018non} and the final formulation is obtained as:
\begin{align*}
 Z_{ij} = \frac{1}{\delta^{'}(X_{ij})}(M_i \odot \text{exp}\{X_{ij}W_{\theta}W_{\psi}^{T}X_{p_{ij}}^{T})\})X_{p_{ij}}W_{g},     \forall i,j ,
\end{align*}
where $\delta^{'}(X_{ij}) = \sum_{p_{ij}} M_i \odot \phi(X_{ij},X_{p_{ij}})$ is the normalization factor, $Z_{ij}$ is the output feature vector at spatial location $i,j$ and $M_i$ is a mask, associated to row $i$, aiming to get rid of the contribution of specific feature vectors in the computation of the new feature vector $Z_{ij}$.  
Considering the receptive field extending upwards, all the pixels in a specific row $i$ are associated with a mask $M_i$ which has weight 1 in row $i$ and all the rows above, and 0 everywhere else. This allows to disregard all Euclidian distances with respect to feature vectors that are not contained in the receptive field extending upwards.
The construction of the mask $M_i$ is not influenced by the shape of the blind-spot structure. The blind-spot shaping always happens right after the four receptive fields are computed, by shifting each of the four feature maps according to the desired final shape.

\begin{table*}
\centering
\caption{Synthetic images - PSNR (dB)}
\label{table:synth_images}
\begin{tabular}{lccccccc}
\textbf{Image} & \textbf{PPB} \cite{5196737} & \textbf{SAR-BM3D} \cite{5989862} & \textbf{Baseline CNN} & \textbf{ID-CNN} & \textbf{Speckle2Void} & \textbf{Speckle2Void + TV} & \textbf{Speckle2Void + NL} \\ \hline \hline
Cameraman & 23.02 & 24.76 & 26.26 & 25.83 & 25.90 & 25.90 & 25.85 \\ \hline
House & 25.51 & 27.55 & 28.17 & 28.32 & 27.96 & 27.94 & 28.08 \\ \hline
Peppers & 23.85 & 24.92 & 26.30 & 26.26 & 25.99 & 26.02 & 26.09 \\ \hline
Starfish & 21.13 & 22.71 & 23.39 & 23.42 & 23.32 & 23.31 & 23.50\\ \hline
Butterfly & 22.76 & 24.48 & 25.96 & 26.09 & 25.82 & 25.80 & 25.98 \\ \hline
Airplane & 21.22 & 22.71 & 23.78 & 23.90 & 23.67 & 23.65 & 23.61\\ \hline
Parrot & 21.88 & 24.17 & 25.91 & 25.85 & 25.44 & 25.45 & 25.46\\ \hline
Lena & 26.64 & 27.85 & 28.66 & 28.71 & 28.54 & 28.58 & 28.44\\ \hline
Barbara & 24.08 & 25.37 & 24.30 & 24.38 & 24.36 & 24.31 & 24.74\\ \hline
Boat & 24.22 & 25.43 & 26.06 & 26.00 & 26.02 & 25.57 & 25.88\\ \hline
\textit{Average} & \textit{23.43} & \textit{24.99} & \textit{25.88} & \textit{25.88} & \textit{25.70} & \textit{25.69} & \textit{25.76}\\ \hline
\end{tabular}%
\end{table*}

\begin{figure*}[t]
  \centering
    \begin{minipage}[b]{\textwidth}
        \begin{minipage}[c]{0.16\textwidth}
        \includegraphics[width=\textwidth]{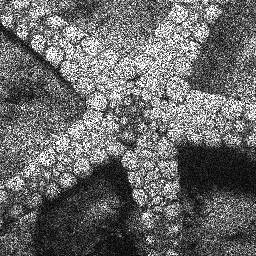}
        \end{minipage}
        \hfill
        \begin{minipage}[c]{0.16\textwidth}
        \includegraphics[width=\textwidth]{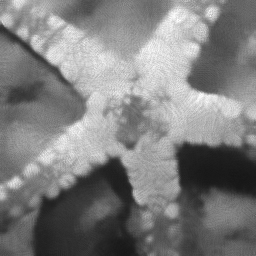}
        \end{minipage}
        \hfill
        \begin{minipage}[c]{0.16\textwidth}
        \includegraphics[width=\textwidth]{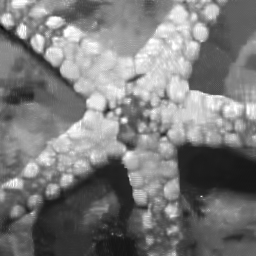}
        \end{minipage}
        \hfill
        \begin{minipage}[c]{0.16\textwidth}
        \includegraphics[width=\textwidth]{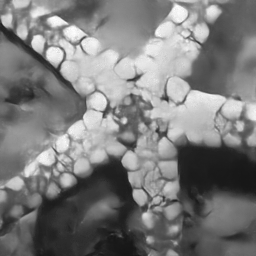}
        \end{minipage}
        \hfill
        \begin{minipage}[c]{0.16\textwidth}
        \includegraphics[width=\textwidth]{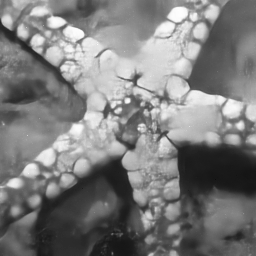}
        \end{minipage}
        \hfill
        \begin{minipage}[c]{0.16\textwidth}
        \includegraphics[width=\textwidth]{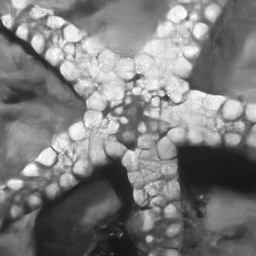}
        \end{minipage}
    \end{minipage}\\
    
    \caption{Synthetic images: Noisy, PPB (21.13 dB), SAR-BM3D (22.71 dB), CNN-based baseline (23.37 dB), ID-CNN (23.42 dB), synthetic Speckle2Void (23.32 dB).}

  \label{fig:zoom_images}
\end{figure*}

\section{Experimental results and discussions}
\label{sec:results}
In this section, we 
evaluate the performance of Speckle2Void, both quantitatively and qualitatively. First, we compare our method with several state-of-the-art methods on a synthetic dataset, where the availability of ground truth images allows to compute objective performance metrics, and then on a real-world SAR dataset, relying on several established no-reference performance metrics and visual results.
Finally, we perform an ablation study to show the impact of various design choices on the despeckling performance. Code is available online\footnote{\url{https://github.com/diegovalsesia/speckle2void}}.

\subsection{Quality assessment criteria}
The evaluation reference metric used to assess quantitative results on synthetic SAR
images corrupted by simulated speckle is the PSNR. This allows to understand the denoising capability of our self-supervised method when compared with traditional methods and CNN-based ones with supervised training.
In the second set of experiments, conducted on real SAR images, we compare the various despeckling methods by relying on some no-reference performance metrics such as equivalent number of looks (ENL), moments of the ratio image ($\mu_r$, $\sigma_r$), quality index $\mathcal{M}$ \cite{Dniz2017UnassistedQE} and the ratio image structuredness RIS \cite{8693546}. 
The ENL is estimated over apparently homogeneous areas in the image and is defined as the ratio of the squared average intensity to the variance. Computing the ENL on the noisy SAR image provides an approximate estimate of its nominal number of looks. Moments of the ratio image $\mu_r$ and $\sigma_r$ measure how close the obtained ratio image is to the statistics of pure speckle ($\mu_r = 1$, $\sigma_r = 1$ are desirable for a single-look image).
The previous metrics lack in conveying information about the detail preservation capability of a filter and the visual inspection of the ratio image would provide an indication of the remaining structure of what ideally should be pure speckle with no visible pattern. To avoid the subjectiveness of the visual interpretation of ratio images, Gomez et al. \cite{Dniz2017UnassistedQE} designed the quality index $\mathcal{M}$.
This index evaluates the goodness of a filter by integrating two measures together: a first-order component measuring the deviation from ideal ENL and from ideal speckle mean over $n$ automatically selected textureless areas and a second-order component measuring the remaining geometrical content within the ratio image through the homogeneity textural descriptor proposed by Haralick et al. \cite{4309314}. Ideally, $\mathcal{M}$ should tend to zero.
RIS \cite{8693546} is a metric closely related to the second-order component of $\mathcal{M}$, allowing to evaluate solely the remaining geometrical content within the ratio image. Similarly to Gomez et al. \cite{Dniz2017UnassistedQE}, it employes the homogeneity textural descriptor proposed by Haralick et al. \cite{4309314} to measure the similarity among neighbouring pixels. RIS is zero when the ratio image consists of independent identically distributed speckle samples.

\subsection{Reference methods}
The following state-of-the-art references are compared with our method on both optical and SAR datasets:
\begin{enumerate}
    \item PPB \cite{5196737};
    \item SAR-BM3D \cite{5989862};
    \item CNN baseline with the improved loss defined in \cite{CozzolinoCNN};
    \item ID-CNN \cite{8053792}.
\end{enumerate}
These methods have been chosen for their popularity and diffusion in the SAR community. For PPB \cite{5196737} and  SAR-BM3D \cite{5989862} methods, we selected parameters as suggested in the original papers.
As a CNN baseline we used the well-known network architecture proposed in \cite{DnCnnZhang}, employing a homomorphic approach and the loss proposed in \cite{CozzolinoCNN} that better adapts to deal with the speckle noise distribution. ID-CNN has been implemented from scratch following the indications in the original paper for what concerns the CNN architecture and the hyperparameters. Notice that both CNNs follow a supervised training approach with synthetically speckled natural images.
We remark that we do not directly compare with the results in SAR-CNN \cite{CozzolinoCNN} or the more recent work in \cite{NonlocalCozzolino} as they use multitemporal data, which would make the setting unfair with respect to the single observation of a scene in our case. In addition, the dataset used in those works is not publicly available.

As described in Sec. \ref{sec:method}, Speckle2Void employs four branches where the horizontal and the vertical directions are processed separately with a different set of parameters, as shown in Fig. \ref{fig:Architecture}. 
The first part of the architecture consists of 17 blocks composed of 2D convolution with $3\times 3$ kernels with 64 filters each, batch normalization and Leaky ReLU nonlinearity. After that, the branches are merged with a series of three $1\times 1$ convolutions.  
The non-local version of our method maintains the same general structure with an addition of 5 non-local layers, one every 3 local layers. 
The same architecture is used in both the experiments with the only difference that in the case of synthetic images the blind-spot shape is $1 \times 1$, since the injected speckle is pixel-wise i.i.d and therefore there is no need to use an enlarged blind-spot. 
Instead, in the real SAR case the blind-spot shape is variable across training. 

For both experiments, the Adam optimization algorithm \cite{kingma2014adam} is employed, with momentum parameters $\beta_{1} = 0.9$, $\beta_{2} = 0.999$, and $\epsilon = 10^{-8}$. 
We use the Tensorflow framework to train the proposed network on a PC with 64-GB RAM, an AMD Threadripper 1920X, and an Nvidia 1080Ti GPU.

\begin{table*}
\centering
\caption{ENL on real SAR test images}
\label{table:real_images}
\begin{tabular}{lccccccc}
\textbf{Metric} & \textbf{Image} & \textbf{PPB} \cite{5196737} & \textbf{SAR-BM3D} \cite{5989862} & \textbf{CNN baseline} & \textbf{ID-CNN} \cite{8053792} & \textbf{Speckle2Void} & \textbf{Speckle2Void NL} \\ \hline \hline
 \multirow{4}{*}{ENL $\uparrow$}  & 1 & 82 & 46.2 &  52.9 & 76.5 & \textbf{88.5} & 86.5\\ 
 \multirow{4}{*}{}    & 2 & 78.6 & 49.1 & 48.7  & 69.9 & \textbf{89.9} & 81.8\\ 
 \multirow{4}{*}{}    & 3 & 76.9 & 58.1 & 52.5 & 73.1 & 84.0 & \textbf{86.0}\\ 
 \multirow{4}{*}{}    & 4 & 54.2 & 40.4 & 37.6 & 46.2 & \textbf{54.7} & 53.1\\ 
 \multirow{4}{*}{}    & 5   & \textbf{22.9} & 16.2 & 14.6 & 16.6 & 18.9 & 17.5\\ \hline
 \multirow{4}{*}{$\mu_r$ $\uparrow$} & 1 & 0.887 & 0.919 &  0.963 & 0.943 & 0.966 & \textbf{0.970}\\ 
 \multirow{4}{*}{}        & 2 & 0.925 & 0.938 & \textbf{0.969} & 0.964 & 0.966 & 0.967\\ 
 \multirow{4}{*}{}        & 3 & 0.926 & 0.941 & \textbf{0.974} & 0.969 & 0.968 & 0.968\\ 
 \multirow{4}{*}{}        & 4 & 0.933 & 0.942 & 0.974 & 0.976 & 0.962 & \textbf{0.977}\\ 
 \multirow{4}{*}{}        & 5 & 0.853 & 0.894 & \textbf{0.950} & 0.918 & 0.947 & 0.946\\ \hline
 \multirow{4}{*}{$\sigma_r$ $\uparrow$} & 1 & \textbf{0.847} & 0.627 &  0.726 & 0.745 & 0.803 & 0.800\\ 
 \multirow{4}{*}{}           & 2 & \textbf{0.886} & 0.674 & 0.740 & 0.803 & 0.829 & 0.817\\ 
 \multirow{4}{*}{}           & 3 & \textbf{0.874} & 0.684 & 0.756 & 0.817 & 0.816 & 0.814\\ 
 \multirow{4}{*}{}           & 4 & \textbf{0.876} & 0.688 & 0.755 & 0.846 & 0.823 & 0.837\\ 
 \multirow{4}{*}{}           & 5 & \textbf{0.891} & 0.549 & 0.683 & 0.664 & 0.748 & 0.736\\ \hline
 \multirow{4}{*}{$\mathcal{M}$ \cite{Dniz2017UnassistedQE} $\downarrow$} & 1 & 24.4 & 16.5 &  11.9 & 14.6 & 7.72 & \textbf{6.71}\\ 
 \multirow{4}{*}{}                                          & 2 & 10.1 & 11.6 & 11.6 & 9.12 & 9.11 & \textbf{8.04}\\ 
 \multirow{4}{*}{}                                          & 3 & 9.82 & 11.3 & 11.3 & 6.93 & 6.24 & \textbf{5.44}\\ 
 \multirow{4}{*}{}                                          & 4 & 10.6 & 10.5 & 12.3 & 9.7 & 8.07 & \textbf{7.74}\\ 
 \multirow{4}{*}{}                                          & 5 & 14.4 & 14.3 & 9.76 & 10.4 & 8.91 & \textbf{7.9}\\ \hline
 
 \multirow{4}{*}{RIS \cite{8693546} $\downarrow$} & 1 & 0.402 & 0.186 &  0.145 & 0.242 & 0.0929 & \textbf{0.0817}\\ 
 \multirow{4}{*}{}    & 2 & 0.114 & 0.0765 & 0.0925 & 0.112 & 0.0918 & \textbf{0.075}\\ 
 \multirow{4}{*}{}    & 3 & 0.114 & 0.0782 & 0.113  & 0.0643 & 0.0396 & \textbf{0.0257}\\ 
 \multirow{4}{*}{}    & 4 & 0.0962 & \textbf{0.0392} & 0.127 & 0.106 & 0.0873 & 0.0804\\ 
 \multirow{4}{*}{}    & 5 & 0.159 & 0.114 & 0.0566 & 0.130 & 0.0708 & \textbf{0.0547}\\ \hline

\end{tabular}%
\end{table*}

\subsection{Synthetic dataset}
\label{sec:synth}

In this experiment we use natural images to construct a synthetic SAR-like dataset. Pairs of noisy and clean images are built by generating i.i.d. speckle to simulate a single-look intensity image ($L=1$).

During training, patches are extracted from 450 different images of the Berkeley Segmentation Dataset (BSD) \cite{MartinFTM01}. The network has been trained for about 400 epochs with a batch size of 16 and learning rate equal to $10^{-5}$.
All the CNN-based methods have been trained with the same synthetic dataset.
Table \ref{table:synth_images} shows performance results on a set of well-known testing images in terms of PSNR.
It can be seen that all the CNN-based methods outperform the non-local traditional methods by a significant margin. Despite ID-CNN employs the suboptimal l2 loss, the TV regularizer helps smoothing out the artifacts, showing approximately the same result as the CNN baseline. 
It can be noticed that our self-supervised method outperforms PPB and SAR-BM3D. Moreover, it is interesting to notice that while the proposed approach does not use the clean data for training, it achieves comparable results with respect to the supervised ID-CNN and CNN-based baseline methods. This happens for the non-local version and TV version as well. We can notice a slight improvement when non-locality is exploited.
Even if we analyze the performance from a qualitative perspective, as done in Fig. \ref{fig:zoom_images}, we observe the same behaviour. Despite the absence of the true clean images during training, our method produces images as visually pleasing as those produced by the CNN-based reference approaches with comparable edge-preservation capabilities.
This is a significant result because it shows that, in theory, we do not need supervised training to achieve the outstanding despeckling results obtained by CNN-based methods.

\begin{figure*}
  \centering
    \begin{minipage}[b]{\textwidth}
        \begin{minipage}[c]{0.32\textwidth}
        \includegraphics[width=\textwidth]{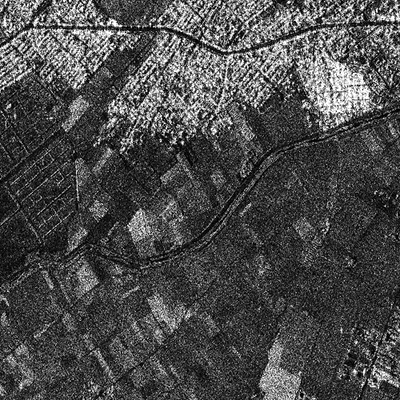}
        \end{minipage}
        \hfill
        \begin{minipage}[c]{0.32\textwidth}
        \includegraphics[width=\textwidth]{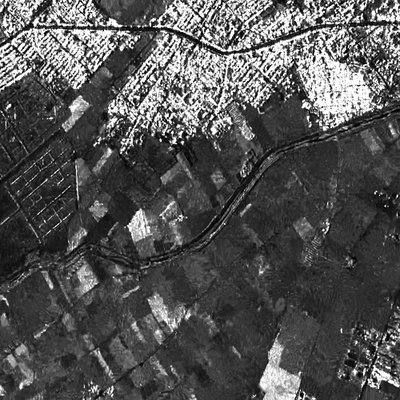}
        \end{minipage}
        \hfill
        \begin{minipage}[c]{0.32\textwidth}
        \includegraphics[width=\textwidth]{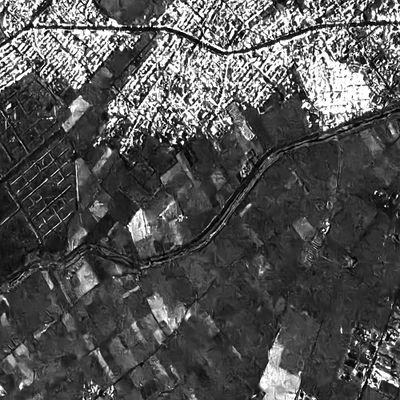}
        \end{minipage}
        \hfill
        
        
        \vspace{0.20cm}
        \begin{minipage}[c]{0.32\textwidth}
        \includegraphics[width=\textwidth]{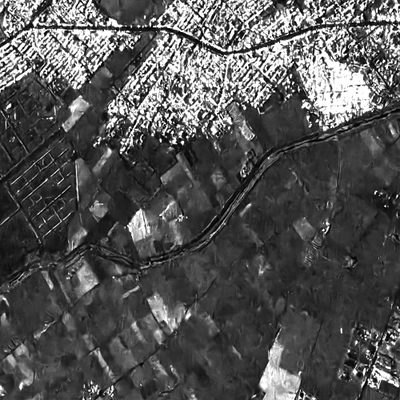}
        \end{minipage}
        \hfill
        \begin{minipage}[c]{0.32\textwidth}
        \includegraphics[width=\textwidth]{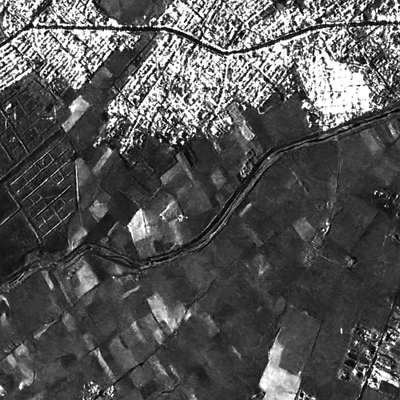}
        \end{minipage}
        \hfill
        \begin{minipage}[c]{0.32\textwidth}
        \includegraphics[width=\textwidth]{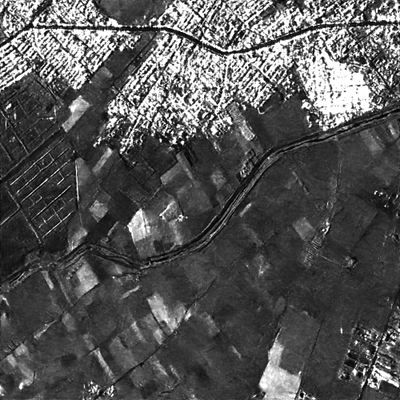}
        \end{minipage}
        \hfill
        
    \end{minipage}\\
    
    \caption{Top-Left to bottom-right: Noisy, SARBM3D, CNN-based baseline, ID-CNN, Speckle2Void, Speckle2Void+NL}
  \label{fig:SAR_image_1}
\end{figure*}

\begin{figure*}[!htb]
  \centering
    \begin{minipage}[b]{\textwidth}
        \begin{minipage}[c]{0.136\textwidth}
        \includegraphics[width=\textwidth]{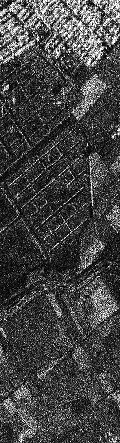}
        \end{minipage}
        \hfill
        \begin{minipage}[c]{0.136\textwidth}
        \includegraphics[width=\textwidth]{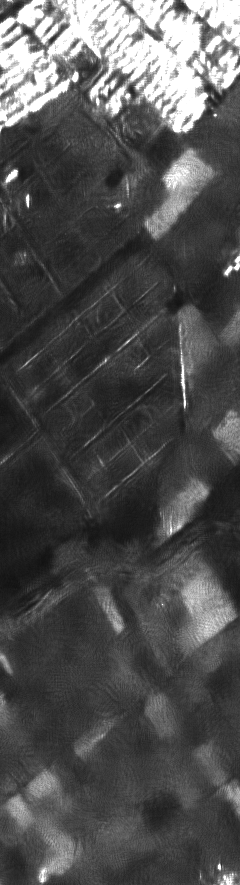}
        \end{minipage}
        \hfill
        \begin{minipage}[c]{0.136\textwidth}
        \includegraphics[width=\textwidth]{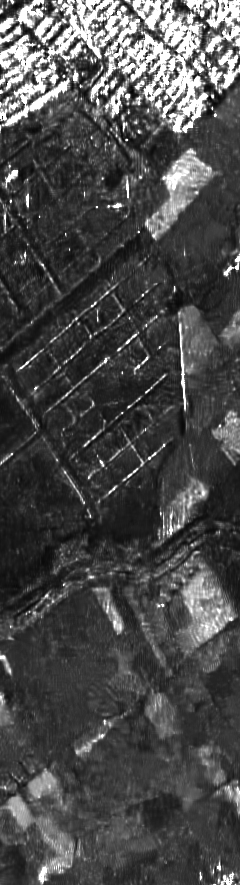}
        \end{minipage}
        \hfill
        \begin{minipage}[c]{0.136\textwidth}
        \includegraphics[width=\textwidth]{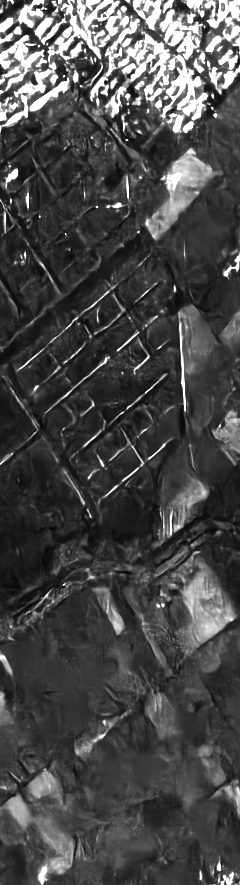}
        \end{minipage}
        \hfill
        \begin{minipage}[c]{0.136\textwidth}
        \includegraphics[width=\textwidth]{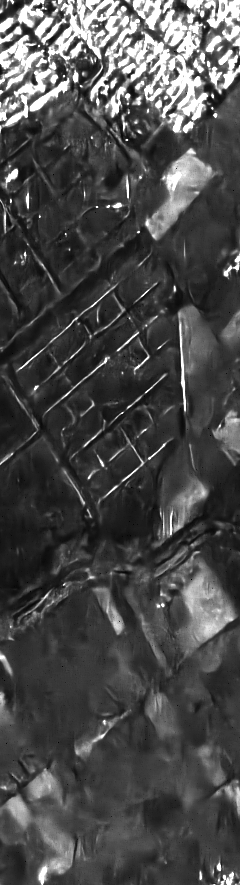}
        \end{minipage}
        \hfill
        \begin{minipage}[c]{0.136\textwidth}
        \includegraphics[width=\textwidth]{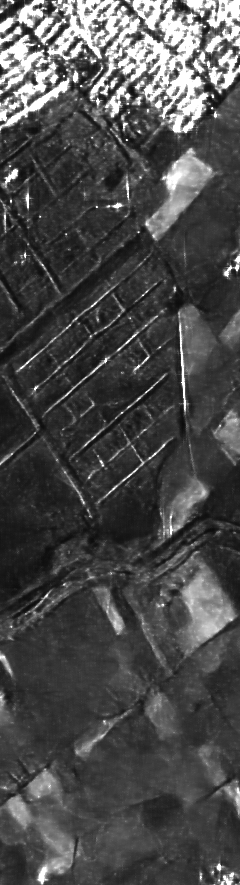}
        \end{minipage}
        \hfill
        \begin{minipage}[c]{0.136\textwidth}
        \includegraphics[width=\textwidth]{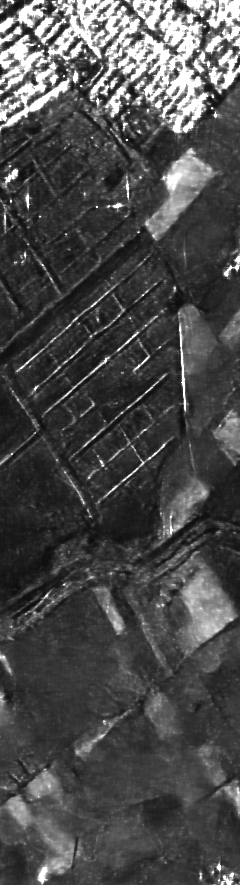}
        \end{minipage}
        \hfill
        
    \end{minipage}\\
    \caption{From left to right: Noisy, PPB, SARBM3D, CNN-based baseline, ID-CNN, Speckle2Void, Speckle2Void+NL}
  \label{fig:slice1}
\end{figure*}

\begin{figure*}[!htb]
  \centering
    \begin{minipage}[b]{\textwidth}
        \begin{minipage}[c]{0.136\textwidth}
        \includegraphics[width=\textwidth]{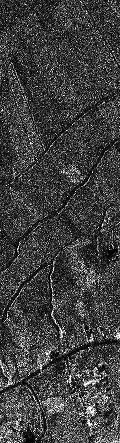}
        \end{minipage}
        \hfill
        \begin{minipage}[c]{0.136\textwidth}
        \includegraphics[width=\textwidth]{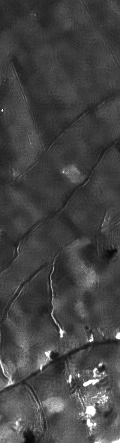}
        \end{minipage}
        \hfill
        \begin{minipage}[c]{0.136\textwidth}
        \includegraphics[width=\textwidth]{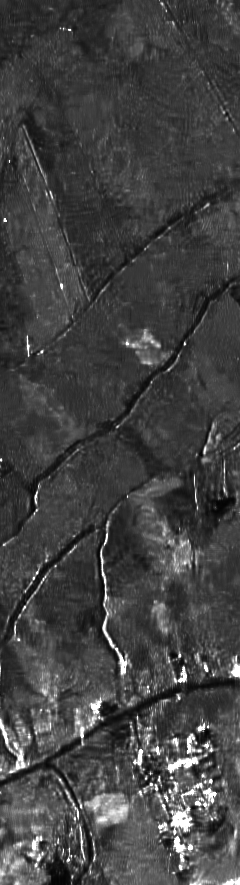}
        \end{minipage}
        \hfill
        \begin{minipage}[c]{0.136\textwidth}
        \includegraphics[width=\textwidth]{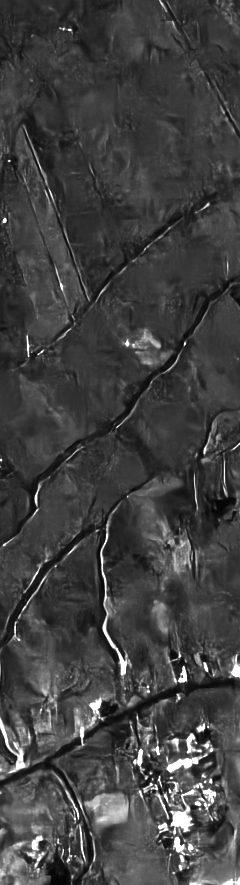}
        \end{minipage}
        \hfill
        \begin{minipage}[c]{0.136\textwidth}
        \includegraphics[width=\textwidth]{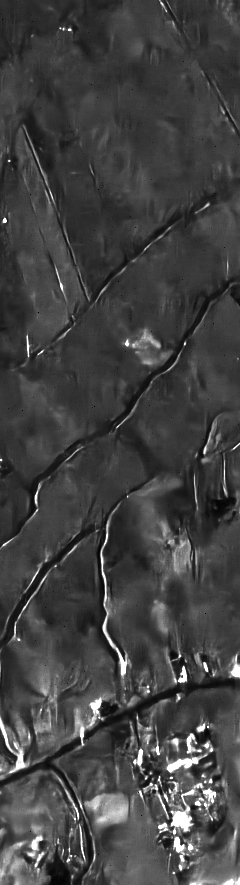}
        \end{minipage}
        \hfill
        \begin{minipage}[c]{0.136\textwidth}
        \includegraphics[width=\textwidth]{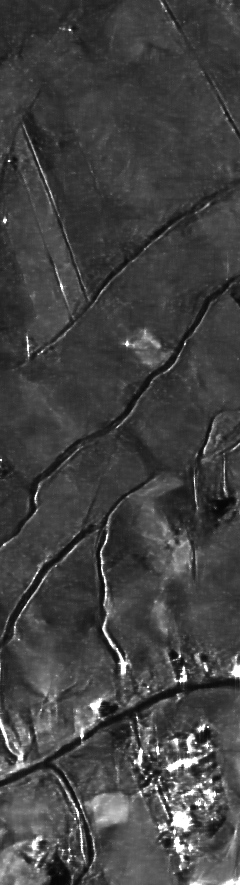}
        \end{minipage}
        \hfill
        \begin{minipage}[c]{0.136\textwidth}
        \includegraphics[width=\textwidth]{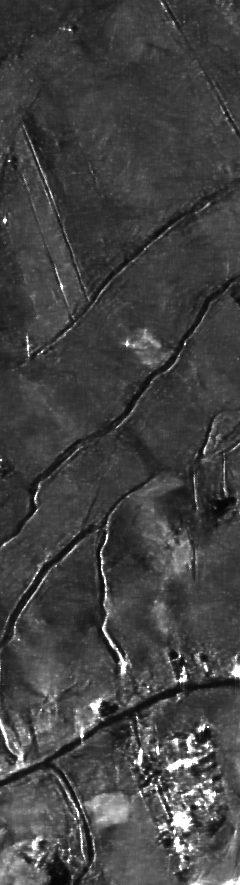}
        \end{minipage}
        \hfill
        
    \end{minipage}\\
    \caption{From left to right: Noisy, PPB, SARBM3D, CNN-based baseline, ID-CNN, Speckle2Void, Speckle2Void+NL}
  \label{fig:slice2}
\end{figure*}

\begin{figure*}[!htb]
  \centering
    \begin{minipage}[b]{\textwidth}
        \begin{minipage}[c]{0.136\textwidth}
        \includegraphics[width=\textwidth]{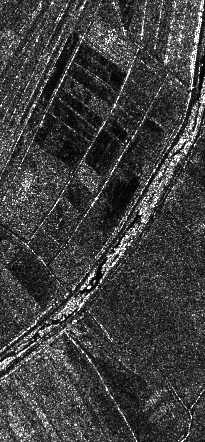}
        \end{minipage}
        \hfill
        \begin{minipage}[c]{0.136\textwidth}
        \includegraphics[width=\textwidth]{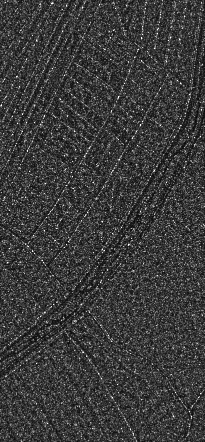}
        \end{minipage}
        \hfill
        \begin{minipage}[c]{0.136\textwidth}
        \includegraphics[width=\textwidth]{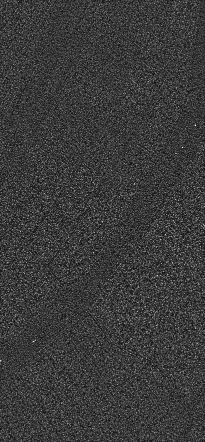}
        \end{minipage}
        \hfill
        \begin{minipage}[c]{0.136\textwidth}
        \includegraphics[width=\textwidth]{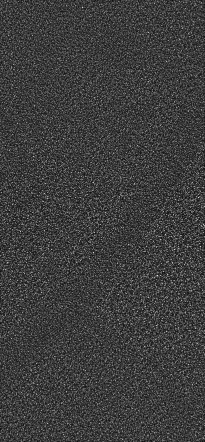}
        \end{minipage}
        \hfill
        \begin{minipage}[c]{0.136\textwidth}
        \includegraphics[width=\textwidth]{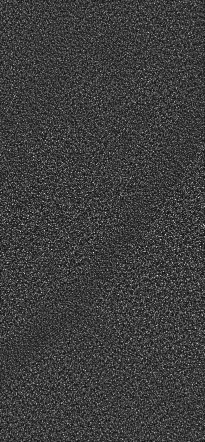}
        \end{minipage}
        \hfill
        \begin{minipage}[c]{0.136\textwidth}
        \includegraphics[width=\textwidth]{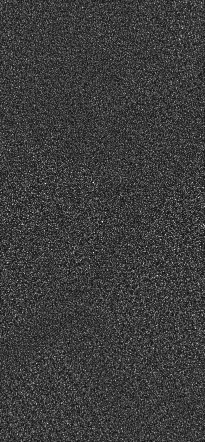}
        \end{minipage}
        \hfill
        \begin{minipage}[c]{0.136\textwidth}
        \includegraphics[width=\textwidth]{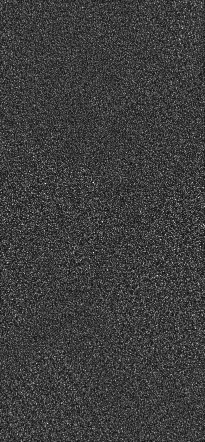}
        \end{minipage}
        \hfill
        
    \end{minipage}\\
    
    \caption{From left to right: Noisy and ratio images (PPB, SARBM3D, CNN-based baseline, ID-CNN, Speckle2Void, Speckle2Void+NL)}
  \label{fig:SAR_ratio}
\end{figure*}

\subsection{TerraSAR-X dataset}
\label{ssec:real}

In this experiment we employ single-look TerraSAR-X images\footnote{\url{https://tpm-ds.eo.esa.int/oads/access/collection/TerraSAR-X/tree}}.
As mentioned in Sec. \ref{sec:training}, 
both training and testing images are pre-processed through the blind speckle decorrelator in \cite{6487399} to whiten them.
To ensure fairness, the whitening procedure is applied to the images for all the tested methods.

During training, $64 \times 64$  patches are extracted from 30000 whitened SAR images of size $256 \times 256$.
The network has been trained for 300000 iterations with a batch size of 16 and an initial learning rate of $10^{-4}$ multiplied by 0.1 at 150000 iterations.
In this case, in addition to two versions (L/NL) of the proposed method used for the synthetic images, we add the TV regularizer to the loss with a $\lambda_{TV}$ of $5\times 10^{-5}$ and we apply the regularized training procedure described in Sec. \ref{subsec:training}, carefully choosing the blind-spot shape. By empirical observation we found non-negligible residual noise correlation in the vertical direction after the whitening stage, so we adapted the structure of the blind spot accordingly. The regularized training alternates between a $3 \times 1$ and $1 \times 1$ shape with probabilities 0.9 and 0.1, respectively. This allows us to take into account the wider vertical autocorrelation of the speckle. In the ablation study presented in Sec. \ref{sec:blindspot_size} we also show the results obtained when only a $1 \times 1$ blind spot is used.

Table \ref{table:real_images} and Figs. \ref{fig:SAR_image_1},\ref{fig:slice1},\ref{fig:slice2} show the results obtained on a set of $1000 \times 1000$ test images\footnote{High-resolution visualization: \url{https://diegovalsesia.github.io/speckle2void}}, that were not included in the training set. 
Speckle2Void outperforms all other methods for almost all testing images in terms of ENL, showing a better speckle suppression capability on smooth areas. The non local version of Speckle2Void scores a slightly lower ENL with respect to the local version as it recovers finer details, generating an additional texture over the apparently homogeneous areas as shown in Fig. \ref{fig:slice1}.
The metric $\mu_r$ is very close to the desired statistic of the ratio image for all the considered methods, in particular for the CNN-based ones. The reference method PPB \cite{5196737} provides the best result in terms of $\sigma_r$ showing a strong speckle suppression, but a very poor detail preservation capability as confirmed by the qualitative comparison in Figs. \ref{fig:slice1} and \ref{fig:slice2}. Despite SAR-BM3D \cite{5989862}  provides worse results in terms of $\sigma_r$ with respect to PPB\cite{5196737}, it produces images with higher fidelity and finer details, as can be observed both visually in Fig. \ref{fig:SAR_image_1} and quantitatively with the RIS \cite{8693546}. 
However, several areas in the SAR-BM3D image still present artifacts like streaks or unrealistic texture.

Overall, the CNN-based methods show a greater speckle suppression than SARBM3D \cite{5989862} and PPB \cite{5196737}. However, both the CNN baseline and ID-CNN \cite{8053792} tend to oversmooth and produce cartoon-like edges. The test image in Fig. \ref{fig:SAR_image_1} presents strong artifacts, making the recovered details look quite unrealistic. This is due to the domain gap between natural images and real SAR images and it represents a strong argument against supervised training with synthetically speckled images. 
On the contrary, Speckle2Void does not hallucinate artifacts over homogeneous regions and produces higher quality images with respect to any other reference method, with much more realistic details in regions with man-made structures and sharp edges. This is confirmed quantitatively by the $\mathcal{M}$ \cite{Dniz2017UnassistedQE} and RIS \cite{8693546} metrics
and qualitatively by a visual inspection of the cleaned image in Fig. \ref{fig:SAR_image_1}, \ref{fig:slice1}, \ref{fig:slice2}.
Instead, Fig. \ref{fig:SAR_ratio} shows the image obtained as the ratio between the noisy and despeckled images. Ideally, no structure should be evident in the ratio image. Also in this case, we can observe the capability of Speckle2Void to remove the speckle effectively, with a minimal amount of visible patterns. The outstanding visual quality of Speckle2Void demonstrates the effectiveness of both direct training on real SAR images and of the adopted regularized training procedure to tackle the residual local noise correlation structure.

Moreover, if we compare the two versions of the proposed method, we can notice that adding the non-local layers provides a marginal improvement in the preservation of the details, yielding lower values for $\mathcal{M}$ \cite{Dniz2017UnassistedQE} and RIS \cite{8693546}. The drawback of the non local version of Speckle2Void is its higher computational overhead, leading to a much longer training time.

\subsection{Ablation study}

\begin{figure}[!htb]
  \centering
    \begin{minipage}[t]{\linewidth}
        \begin{minipage}[t]{0.32\linewidth}
        \includegraphics[width=\linewidth]{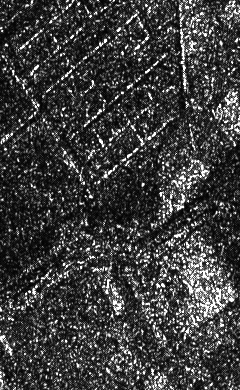}
        \end{minipage}
        \hfill
        \begin{minipage}[t]{0.32\linewidth}
        \includegraphics[width=\linewidth]{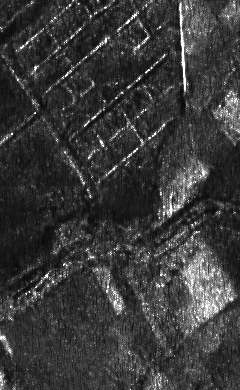}
        \end{minipage}
        \hfill
        \begin{minipage}[t]{0.32\linewidth}
        \includegraphics[width=\linewidth]{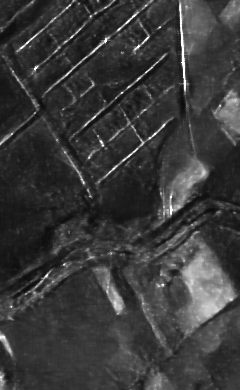}
        \end{minipage}
        \hfill
        
        \vspace{0.10cm}
        \begin{minipage}[t]{0.32\linewidth}
        \includegraphics[width=\linewidth]{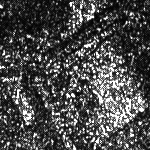}
        \end{minipage}
        \hfill
        \begin{minipage}[t]{0.32\linewidth}
        \includegraphics[width=\linewidth]{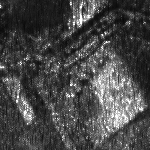}
        \end{minipage}
        \hfill
        \begin{minipage}[t]{0.32\linewidth}
        \includegraphics[width=\linewidth]{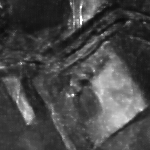}
        \end{minipage}
        \hfill
        
    \end{minipage}\\
    
    \caption{From left to right: cleaned image resulting from the training with the original TerraSAR-X dataset (ENL 1.28), cleaned image resulting from the training with the whitened TerraSAR-X dataset (ENL 14.5) and Speckle2Void (ENL 88.5).}
  \label{fig:white_nowhite}
\end{figure}

In the following study, we want to assess the benefits of some of the features proposed for Speckle2Void.

\subsubsection{Original vs whitened}\label{sec:blindspot_size}
First, we show the importance of the pixel-wise noise independence condition when training a blind-spot network. To assess it, we train Spleckle2Void with two different datasets. One dataset is composed of real single-look complex images as they are provided by the focusing algorithm for the TerraSAR-X satellite, while the other dataset is composed of the same real SAR images but pre-processed by the decorrelator defined in \cite{6487399}. For both datasets we use a $1\times1$ blind-spot shape, including solely the center pixel during the entire training. To better highlight the effect of the whitening procedure, we do not add the TV regularization in the loss.
Fig. \ref{fig:white_nowhite} shows the two resulting cleaned images together with the one obtained by the full Speckle2Void method (whitening+variable blind spot). The visual difference between the left image and the middle one shows that the decorrelator improves drastically the qualitative performance, since barely any denoising is performed in the first image. 

\subsubsection{Enlarging the blind-spot}
\begin{figure}[t]
  \centering
    \begin{minipage}[t]{\linewidth}
        \begin{minipage}[t]{0.32\linewidth}
        \includegraphics[trim={0cm 0cm 0cm 
        0cm},clip,width=\linewidth]{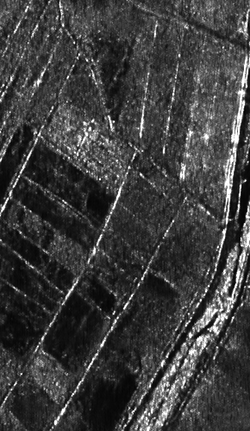}
        \end{minipage}
        \hfill
        \begin{minipage}[t]{0.32\linewidth}
        \includegraphics[trim={0cm 0cm 0cm 
        0cm},clip,width=\linewidth]{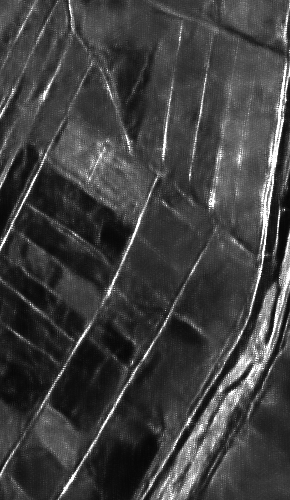}
        \end{minipage}
        \hfill
        \begin{minipage}[t]{0.32\linewidth}
        \includegraphics[trim={0cm 0cm 0cm 
        0cm},clip,width=\linewidth]{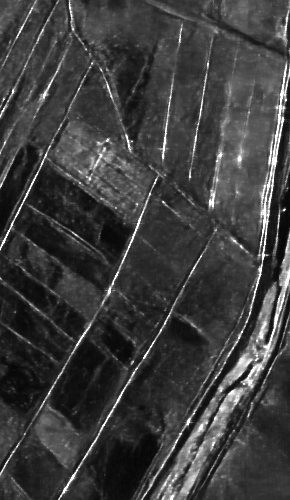}
        \end{minipage}
        \hfill
         
        \vspace{0.10cm}
        \begin{minipage}[t]{0.32\linewidth}
        \includegraphics[width=\linewidth]{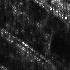}
        \end{minipage}
        \hfill
        \begin{minipage}[t]{0.32\linewidth}
        \includegraphics[width=\linewidth]{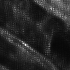}
        \end{minipage}
        \hfill
        \begin{minipage}[t]{0.32\linewidth}
        \includegraphics[width=\linewidth]{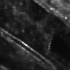}
        \end{minipage}
        \hfill
        
    \end{minipage}\\
    
    \caption{From left to right: network with $1\times1$ blind-spot, network with $3\times3$ blind-spot, Speckle2Void}
  \label{fig:larger_blindspot}
\end{figure}

In our regularized training procedure we vary the shape of the blind-spot to account for the residual noise correlation that persists even after the whitening procedure. To better understand the effect of enlarging the size of the blind-spot structure, we compare Speckle2Void trained with the canonical $1 \times 1$ blind-spot shape against a $3 \times 3$ shape. Notice that, in this experiment, the latter uses the $3 \times 3$ blind-spot in testing as well, differently from the regularization procedure explained in \ref{sec:training} which always uses a $1 \times 1$ blind spot in testing. Moreover, to better highlight the effect of the shape of the blind-spot, we do not add the TV regularization in the loss.
Fig. \ref{fig:larger_blindspot} shows a visual comparison between the two methods. The left image is the result produced by the network with blind-spot of shape $1 \times 1$. We can notice sharper edges and more details with respect to the middle image produced by the network with blind-spot of shape $3 \times 3$, which looks more blurry. However, we also see more residual noise in the image on the left. Enlarging the shape of blind-spot structure leads to a more effective speckle noise reduction as the network uses surrounding pixels that are less correlated with center pixel. A downside of expanding the blind-spot is to reduce the amount of relevant information for the network to estimate the center pixel, resulting in a smoother image with a loss of high frequency details, failing to preserve the original edges. In the image on the right we report the result of Speckle2Void, showing that the proposed method is able to achieve stronger speckle suppression with an impressive preservation of details.

\subsubsection{Effect of the TV regularizer}
\begin{figure}[t]
  \centering
    \begin{minipage}[t]{\linewidth}
        \begin{minipage}[t]{0.32\linewidth}
        \includegraphics[width=\linewidth]{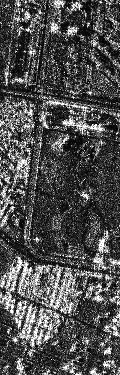}
        \end{minipage}
        \hfill
        \begin{minipage}[t]{0.32\linewidth}
        \includegraphics[width=\linewidth]{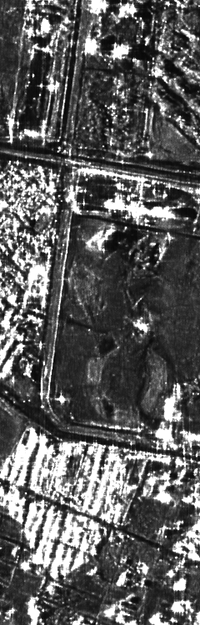}
        \end{minipage}
        \hfill
        \begin{minipage}[t]{0.32\linewidth}
        \includegraphics[width=\linewidth]{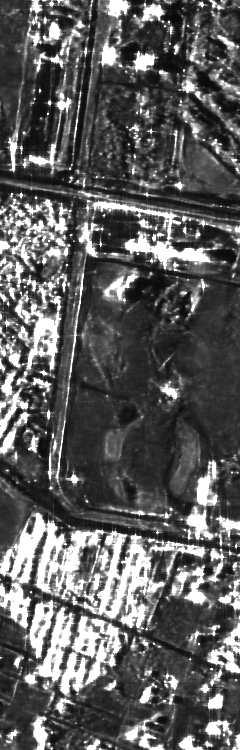}
        \end{minipage}
        \hfill
        
        \vspace{0.10cm}
        \begin{minipage}[t]{0.32\linewidth}
        \includegraphics[width=\linewidth]{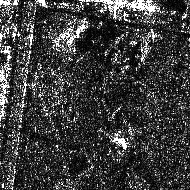}
        \end{minipage}
        \hfill
        \begin{minipage}[t]{0.32\linewidth}
        \includegraphics[width=\linewidth]{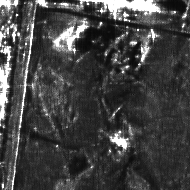}
        \end{minipage}
        \hfill
        \begin{minipage}[t]{0.32\linewidth}
        \includegraphics[width=\linewidth]{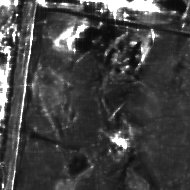}
        \end{minipage}
        \hfill
        
    \end{minipage}\\
    
    \caption{From left to right: Noisy, Speckle2Void w/o TV and Speckle2Void.}
  \label{fig:proposed_NOTV}
\end{figure}

Speckle2Void employs TV in the loss as an additional spatial regularizer. We aim to understand its impact by comparing Speckle2Void with a version trained without TV. Fig. \ref{fig:proposed_NOTV} shows the resulting cleaned images, revealing the reduced amount of artifacts and smoother flat areas when the TV regularization is employed. 

\subsubsection{Prior vs posterior}
\begin{figure}[t]
  \centering
    \begin{minipage}[t]{\linewidth}
        \begin{minipage}[t]{0.32\linewidth}
        \includegraphics[width=\linewidth]{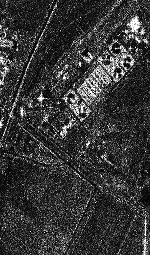}
        \end{minipage}
        \hfill
        \begin{minipage}[t]{0.32\linewidth}
        \includegraphics[width=\linewidth]{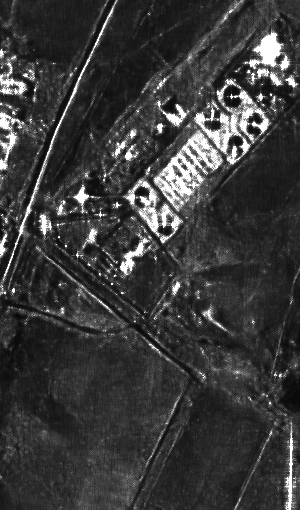}
        \end{minipage}
        \hfill
        \begin{minipage}[t]{0.32\linewidth}
        \includegraphics[width=\linewidth]{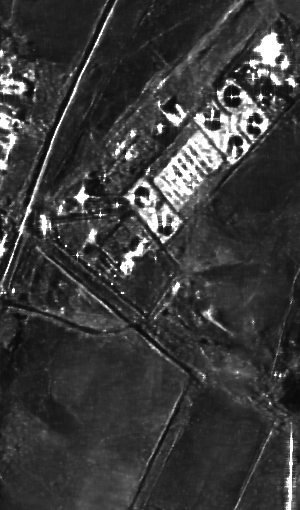}
        \end{minipage}
        \hfill
        
        \vspace{0.10cm}
        \begin{minipage}[t]{0.32\linewidth}
        \includegraphics[width=\linewidth]{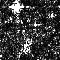}
        \end{minipage}
        \hfill
        \begin{minipage}[t]{0.32\linewidth}
        \includegraphics[width=\linewidth]{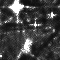}
        \end{minipage}
        \hfill
        \begin{minipage}[t]{0.32\linewidth}
        \includegraphics[width=\linewidth]{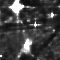}
        \end{minipage}
        \hfill
        
    \end{minipage}\\
    
    \caption{From left to right: Noisy, Speckle2Void (Prior mean image), Speckle2Void (Posterior mean image).}
  \label{fig:prior_post}
\end{figure}

The Bayesian framework, exploited in our method, makes use of the noisy SAR image to obtain the despeckled version by computing the expected value of the posterior distribution. The blind-spot CNN produces the parameters of the prior distribution. If we compute its expected value we obtain the prior despeckled image. In Fig. \ref{fig:prior_post}, the prior and the posterior images highlight the great qualitative improvement brought by the use of the noisy observations in the estimation of the cleaned image with the posterior mean. The prior image shows fuzzy edges and a disturbing granular pattern that makes the posterior image visually preferable.

\section{Conclusion}
\label{sec:conclusions}

In this paper we have presented Speckle2Void, a self-supervised Bayesian denoising framework for despeckling. The main obstacle in applying classical supervised deep learning methods to despeckling is represented by the vast content disparity between speckle injected natural images and real SAR images, often resulting in unfaithful cleaned images. Speckle2Void exploits a customized version of the blind-spot convolutional networks where the receptive field is constrained to exclude a variable amount of pixels throughout training to account for the correlation structure of the noise, introducing one of the first deep learning despeckling method purely based on real single-look complex SAR images. Speckle2Void is able to learn to produce excellent images with faithful details and no visible residual speckle noise.

\bibliographystyle{IEEEtran}
\bibliography{biblio}

%








\end{document}